\documentclass[printer]{aa} 

\usepackage{graphicx}
\usepackage{txfonts}
\usepackage{natbib,twoopt}

\usepackage{enumitem}

\usepackage[breaklinks=true]{hyperref} 
\bibpunct{(}{)}{;}{a}{}{,}             
\makeatletter
  \newcommandtwoopt{\citeads}[3][][]{\href{http://adsabs.harvard.edu/abs/#3}%
    {\def\hyper@linkstart##1##2{}%
     \let\hyper@linkend\@empty\citealp[#1][#2]{#3}}}
  \newcommandtwoopt{\citepads}[3][][]{\href{http://adsabs.harvard.edu/abs/#3}%
    {\def\hyper@linkstart##1##2{}%
     \let\hyper@linkend\@empty\citep[#1][#2]{#3}}}
  \newcommandtwoopt{\citetads}[3][][]{\href{http://adsabs.harvard.edu/abs/#3}%
    {\def\hyper@linkstart##1##2{}%
     \let\hyper@linkend\@empty\citet[#1][#2]{#3}}}
  \newcommandtwoopt{\citeyearads}[3][][]%
    {\href{http://adsabs.harvard.edu/abs/#3}
    {\def\hyper@linkstart##1##2{}%
     \let\hyper@linkend\@empty\citeyear[#1][#2]{#3}}}
\makeatother

\begin{document}

\title{Volume density maps of the 862~nm DIB carrier and interstellar dust}
\subtitle{A hint for the role of carbon-rich ejecta from AGB stars?}

\author{N.~L.~J.~Cox \inst{1}
      \and
      J.~L.~Vergely \inst{2}
      \and
      R.~Lallement \inst{3}          
      }

\institute{
   ACRI-ST, Centre d’Etudes et de Recherche de Grasse (CERGA), 10 Av. Nicolas Copernic, 06130 Grasse, France \email{nick.cox@acri-st.fr}
\and
    ACRI-ST, 260 Route du Pin Montard, Sophia-Antipolis, France \email{jean-luc.vergely@acri-st.fr}
\and
    GEPI, Observatoire de Paris, Université PSL, CNRS, 5 place Jules Janssen, 92195 Meudon, France \email{rosine.lallement@obsmp.fr}
}

\date{Received 10 April 2024}

  \abstract
   {The carbonaceous macromolecules imprinting in astronomical spectra the numerous absorptions called “Diffuse Interstellar Bands” (DIBs) are omnipresent in the Galaxy and beyond and represent a considerable reservoir of organic matter. However, their chemical formulae, formation and destruction sites remain open questions. Their spatial distribution and the local relation to other interstellar species is paramount to unravel their role in the lifecycle of organic matter.}
   {Volume density maps bring local instead of line-of-sight distributed information, and allow for new diagnostics. We present the first large-scale volume (3D) density map of a DIB carrier and compare it with an equivalent map of interstellar dust.}
   {The DIB carrier map is obtained through hierarchical inversion of $\sim$202,000 measurements of the 8621~nm DIB obtained with the Gaia-RVS instrument. It covers about 4000 pc around the Sun in the Galactic plane. A dedicated interstellar dust map is built based on extinction towards the same target stars.}
   {At the $\simeq$~50 pc resolution of the maps, the 3D DIB distribution is found remarkably similar in shape to the 3D distribution of dust. On the other hand, the DIB-to-dust local density ratio increases in low-dust areas. It is also increasing away from the disk, however, the minimum ratio is found to be shifted above the Plane to Z=$\simeq$~+50pc. Finally, the average ratio is also surprisingly found to increase away from the Galactic Center. We suggest that the three latter trends may be indications of a dominant contribution of material from the carbon-rich category of dying giant stars to the formation of the carriers.  Our suggestion is based on recent catalogues of AGB stars and estimates of their mass fluxes of C-rich and O-rich ejecta.}
   {}

   \keywords{Interstellar medium (ISM) --
             dust, extinction --
             ISM: lines and bands --
             ISM: molecules --
             ISM: structure 
               }

   \titlerunning{Volume density maps of the 862~nm DIB carrier and interstellar dust}

   \maketitle

\section{Introduction}

Diffuse interstellar bands (DIBs) are absorptions of various shapes and depths distributed in the optical and near-infrared spectra of distant objects. DIB catalogues have been constantly enlarged due to the accumulation of stellar spectra of increasing quality, reaching more than 500 different features (see \citealt{2019ApJ...878..151F} and references therein). DIBs are omnipresent in the Universe, both detected in the local Galaxy as well as in neighbouring and distant galaxies \citep{2000ApJ...537..690H,2011ApJ...726...39C,2015A&A...576L...3M}. The column densities of their carriers (i.e., the absorbing species), measured through the fraction of absorbed light, are long known to correlate, to varying degree, with the amount of interstellar dust or gas \citep{2011A&A...533A.129V,2011ApJ...727...33F}, making them potential proxies for the columns of foreground interstellar matter \citep{2013ApJ...779...38P}. Detailed analysis of the band spectral profiles reveals structures that are indicative of rotating large molecules \citep{1995MNRAS.277L..41S} whose presence in the Interstellar Medium (ISM) is also known from their infrared signatures arising from different stretching/vibration modes of polycyclic aromatic hydrocarbons \citep{1985A&A...146...81L}. However, despite progress in laboratory astrophysics, their carriers remain, in effect, unidentified \citep{2014IAUS..297.....C}, at the exception of only four bands, situated in the near-infrared part of the spectrum which have been assigned to the ionized buckminsterfullerene C$_{60^+}$ \citep{1994Natur.369..296F,2015Natur.523..322C,2019ApJ...875L..28C}.

Adding to the mystery of their composition, and despite all measurements, there is no consensus on their formation sites and processes, nor on the way they are destroyed or transformed. The main reason is the poor diagnostic on the location of the carrier provided by individual absorption measurements; indeed, the location can be anywhere between the Sun and the observed target star. This is why, besides identifications of the carriers that would undoubtedly be an important discovery in terms of understanding the chemical complexity and composition of organic matter in space, constraining their spatial distribution is equally important since it will shed light on their formation and destruction sites and, in turn, on the processes at play and the effects of local physical conditions. Once we know the constraints on their formation and lifetimes, and even without identifying them, DIBs can be used as probes of the physico-chemical conditions of the interstellar medium, not only in the local galaxy, but also further away in in distant galaxies, permitting unique diagnostics.

Globally, all DIBs are found to be positively correlated with the amount of dust and total gas columns. However, there is a large scatter on these relations and the dependence on molecular hydrogen columns may be negative, suggesting environmental conditions play a fundamental role in the relative abundance of the carriers (e.g. \citealt{2011ApJ...727...33F}). Indeed, according to the most detailed studies of nearby dense clouds, the strongest DIBs appear to respond differently to changes in the local physical conditions of ISM \citep{2019ApJ...878..151F,2011A&A...533A.129V,2011ApJ...727...33F}. In particular, the local effective radiation field is affecting the relative abundances between different DIB carriers \citep{2011A&A...533A.129V}. Also, most DIB carriers become depleted in opaque regions, i.e., towards the interior of dense interstellar clouds. This effect is known for a long time \citep{1974ApJ...194..313S} and has now been quantitatively documented in various ways \citep{2015MNRAS.452.3629L,2017ApJ...850..194F,2017A&A...605L..10E}. Furthermore, the ratio between DIB carrier column and reddening starts to decrease for lines-of-sight crossing the more central regions of dense clouds, at an average reddening E(B-V) = 0.5 -- 1.0~mag \citep{2015MNRAS.452.3629L}, or at an average molecular fraction of 0.3 \citep{2017ApJ...850..194F}. This is called the ``skin effect''. However, other factors enter in play as suggested from the principle-component analysis presented by \citet{2017ApJ...836..162E}. 

The 862~nm DIB is a broad band (Full Width at Half Maximum $\sim$ 0.4 nm), whose relatively small depth makes it challenging to measure in the spectra of cool stars. However, being spectrally close to ionized calcium lines commonly used to constrain the stellar parameters, its study benefits from stellar spectroscopic surveys using these stellar lines. Consequently, its properties have been well studied, especially thanks to the ground-based stellar spectroscopic survey RAVE \citep{2013ApJ...778...86K}. They have been found to be similar to the majority of bands, and the correlation between the 862~nm DIB and interstellar reddening (or visual extinction) has been found to be relatively tight \citep{2013ApJ...778...86K,2014Sci...345..791K,2015A&A...573A..35P}. The measurements did not allow yet to detect the depletion in very opaque regions. Its study was recently boosted due to its presence in the spectral interval selected for the Radial Velocity Spectrometer (RVS), the high-resolution spectrometer on board Gaia. With Gaia, DIB measurements could be coupled with high precision parallaxes of target stars. \citet{2023A&A...674A..40G} presented individual DIB measurements from Gaia Data Release 3 for 476,000 stars based on a comparison of observed spectra with stellar models and explored the relationship with the extinction and the target location.

Maps of DIB equivalent widths (EWs) or absorption strengths, quantities proportional to integrated DIB carrier density columns, have been presented for several DIBs based on ground surveys \citep{2014Sci...345..791K,2015MNRAS.447..545B,2015ApJ...798...35Z} and more recently on Gaia data \citep{2023A&A...674A..40G,2022yCat.1355....0G,2023A&A...680A..38G}. They display for each location in space the measured cumulative value of DIB carriers, integrated between the Sun and the target star. All these pseudo-3D maps revealed a similarity with the reddening, but there are some apparently contradictory results. Based on integrated line-of-sight values of the 862~nm DIB observed with RAVE, \citet{2014Sci...345..791K} found a DIB scale height above the Galactic disk of about 200~pc, far above the dust scale height of about 100~pc, while \citet{2023A&A...674A..40G} derived for the same DIB and from Gaia data a scale height similar to the one of dust. More precisely, using full-sky Gaia data, the authors found evidence for a significant variability of the scale height according to the Galactic area, and attributed the difference with respect to the RAVE result to the use of different longitude and latitude ranges for the two studies.

At variance with mapping of Sun-star integrated quantities, full-3D tomography aims at computing the local volume density of a species at each point in space. A 3D mapping of the strong 578~nm DIB has been presented by \citet{2019NatAs...3..922F} for the so-called Local Bubble, a 250~pc wide region around the Sun, based on 360 lines of sight. Nowadays, Milky Way large-scale full-3D maps have been constructed solely for interstellar dust, using extinction/colour excess measurements of Galactic target stars. To do so, differentials of the cumulative data along radial directions \citep{2019MNRAS.483.4277C,2019ApJ...887...93G} or full-3D inversion techniques \citep{2001A&A...366.1016V,2020A&A...639A.138L,2020A&A...643A.151R} were applied to a whole set of integrated data. During the last years, 3D extinction maps have been constantly improved, following the accurate parallaxes and photometric data from Gaia \citep{2022A&A...661A.147L,2022A&A...664A.174V}. Here we build the first, large-scale, full-3D map of a DIB carrier, tracing it out to $\sim$2000~pc from the Sun, and do the first comparisons between local values of DIB carriers and dust grains. 

The data are presented in Sect.~\ref{sec:data} and the methodology in Sect.~\ref{sec:method}. The results are presented in Sect.~\ref{sec:results} and discussed in Sect.~\ref{sec:discussion}.

\section{Data}\label{sec:data}

This study takes as input the Gaia DR3 data release \citep{2022yCat.1355....0G} which includes DIB measurements for 463,486 sources, constituting the largest all-sky survey of DIBs to date \citep{2023A&A...674A..40G}. The reported DIB equivalent width values are integrated line-of-sight measurements.
%
%
The following procedure, considering the recommendations from \citet{2023A&A...674A..40G}, is defined to prepare a high-fidelity input catalogue for inversion:

\begin{enumerate}
    \item Select all (463,486) objects with reported DIB equivalent widths in Gaia DR3 
    \item From step 1, select all objects that have:
        \begin{itemize}
            \item[$\bullet$] parallax error $<$ 20~\% (ePlx/Plx $< 0.2$)
            \item[$\bullet$] DIB quality flag = 0, 1, or 2 (qDIB $< 3$)
            \item[$\bullet$] |z| $<$ 400 pc (only sources within 400~pc of the galactic plane)
            \item[$\bullet$] $r < 5000$~pc (only sources within 5~kpc of the Sun)
        \end{itemize}
\end{enumerate}

\noindent This produces an input catalogue of 202,340 sources. We further investigated if additional filtering benefits the reconstruction. \citet{2023ApJ...954..141S} argue that residual stellar lines not considered by the Gaia processing pipeline are mistakenly fitted as DIBs. The  authors produced a catalogue of cleaned measurements based on their own analysis of publicly available RVS spectra, however, due to the limited number of such public data, this catalogue is not large enough to be used for inversion.  We attempted to exclude features at 861.6, 861.7, 862.5~nm which are thought to be stellar lines masquerading as DIBs. Note that $\lambda$DIB is measured in the stellar rest frame, to which real DIBs would not be expected to align. Therefore, to remove the most obvious stellar line contamination, we apply the following filtering on $\lambda_\mathrm{DIB}$ (in nm):

\begin{itemize}[leftmargin=10mm]
    \item[$\bullet$] $862.52 < \lambda_\mathrm{DIB} < 862.58$
    \item[$\bullet$] $861.76 < \lambda_\mathrm{DIB} < 861.78$
    \item[$\bullet$] $861.66 < \lambda_\mathrm{DIB} < 861.70$
\end{itemize}

\noindent This reduces the input catalogue size to 175,578 sources. However, the effect of this filtering (tested also with applying variations on the exact lambda boundaries) on the map reconstruction turned out to be negligible. 
Another effect we noticed in the input catalogue is that the relation between DIB strength and distance changes with the effective temperature of the background star for targets at small distances. This suggests a systematic uncertainty in the measurement of the DIB strength most noticeable at shortest distances where the DIBs are weakest. Because it essentially affects the distribution very close to the Sun, we did not remove those targets. 

Finally, additional filtering could also be applied to the DIB EW uncertainty (eEWDIB). We examined filtering on eEWDIB/EWDIB $<$ 0.3, but it removes too much data, particularly for the low EW values (the input catalogue is reduced to 122,585 sources).

In fact, the reconstruction algorithm manages errors and outliers in a robust way. During the tomography steps we do not use individual DIB equivalent widths, but an average of many values made over spatial boxes. The size of the boxes depends on the resolution. For a 100-pc resolution reconstruction, boxes have a size of 100x100x100 pc. The algorithm averages each DIB equivalent width by 1/eEWDIB$^2$. During the inversion, the average EWDIB is used in each box together with an estimation of the error of the average in the box. The error in each box is computed from the standard deviation of the individual EWDIBs present in the box. Incorrect EWDIB values in a box will lead to an increase of the standard deviation and the data (average over the box) will have lower weight in the result. Thus, pre-filtering on eEWDIB does not significantly alter (improve) the results. Similarly, the algorithm is robust against biases related to the effective temperature of the background stars.
Following the adage “less is more” the simpler source filter (“step 2” above) is adopted.

\begin{figure}
\centering 
\includegraphics[width=\columnwidth]{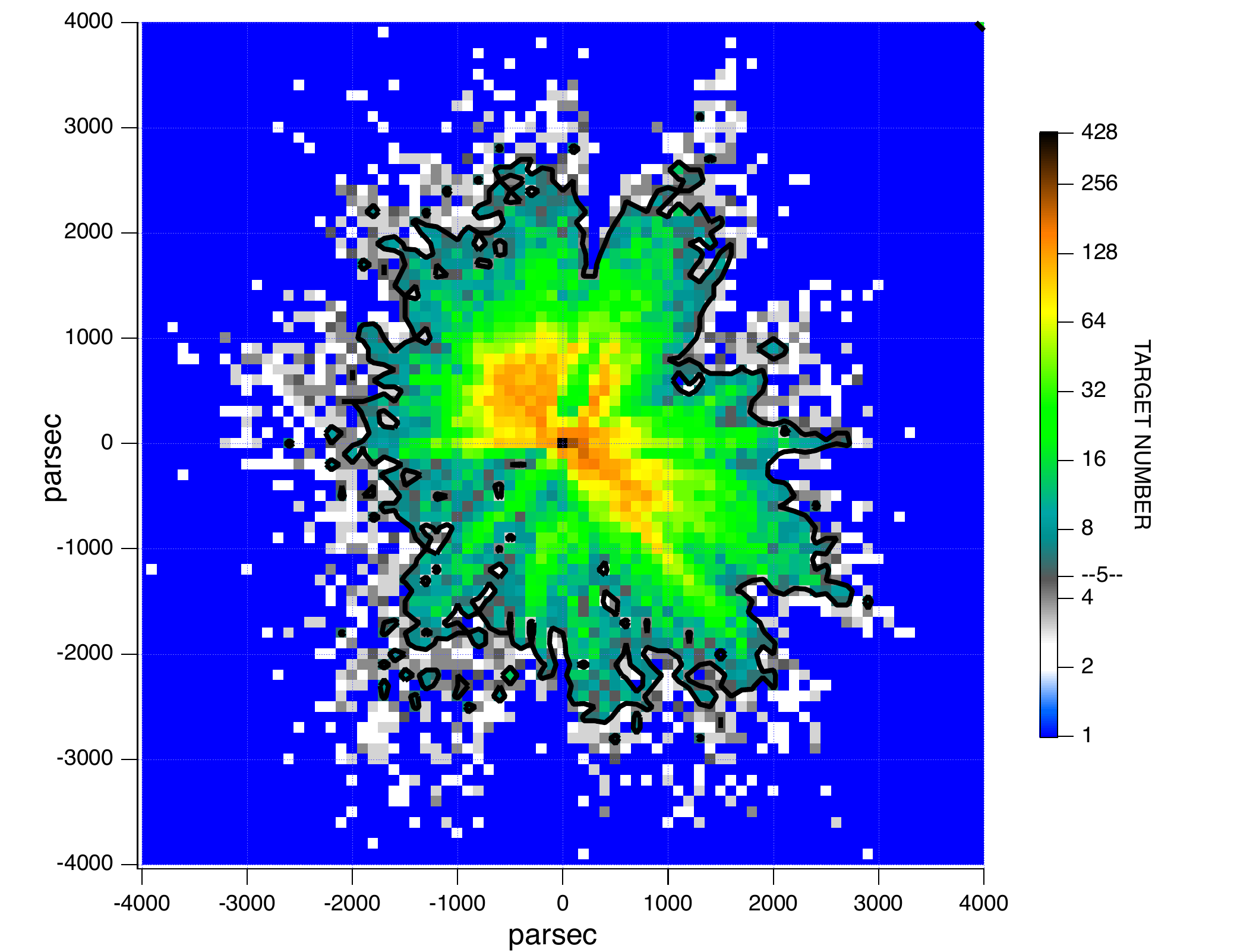}
\caption{Coverage of sources in the input catalogue along the Galactic plane. The number of sources per 100$^{3}$ pc$^{3}$ cell is color coded. All sources with abs($z$) $<$ 50~pc are included. The limit of 5 targets per cell, used in the statistical study, is marked by a thick black line. }
\label{fig:5}
\end{figure}

\begin{figure}
\centering 
\includegraphics[width=\columnwidth]{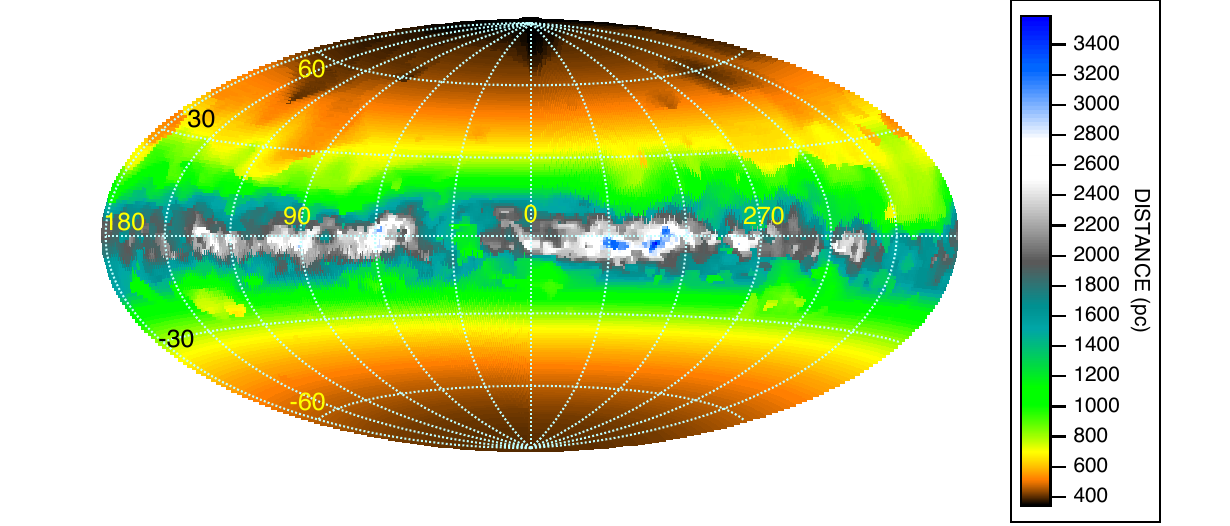}
\caption{Aitoff projection of the maximum reliable distance map for the DIB 3d reconstruction. Iso-Galactic longitudes and latitudes are drawn spaced by 30\degr. The Galactic Center ($l=0$) is at the center and longitudes are increasing to the left. The reconstruction in the Galactic plane (|b| $<$ 15\degr) has the highest fidelity.}
\label{fig:6}
\end{figure}

\section{Methodology}\label{sec:method}

\subsection{Tridimensional (3D) tomography}

The reconstruction of the volume (3D) density of either interstellar dust grains producing the extinction or of DIB carriers is based on the inversion of a catalogue of measured integrated values, i.e. as seen from the origin up to a given point (x,y,z) at distance d, the target star location. The integrated quantity is the stellar light extinction in the first case, and the absorption equivalent width in the second. However, this catalogue is made of a limited and irregular sampling of locations in 3D space. Moreover, the accuracy of this reconstruction is limited by the accuracy of the distance d, the uncertainty on the measured quantity, the sampling density, and the choice of the prior distribution. As a result, the spatial resolution and quality of the final density maps derived through inversion depends in large part on the number of data points, and their accuracy, that sample the true 3D distribution. Increasing the number of data points will improve the retrieval of spatial information, particularly in the radial direction. On the other hand, allowing the inclusion of inaccurate data points will lead to the appearance of false features. Consequently, filtering applied in Sect.~\ref{sec:data} is essential.

We applied to the selected data (cf. Sect.~\ref{sec:data}) the same hierarchical, iterative inversion technique used in \citet{2022A&A...661A.147L} and \citealt{2022A&A...664A.174V} for the dust extinction. According to this technique, the achievable spatial resolution of the map is variable in 3D space and governed by the target star density at each location. At each iteration the maximum resolution (or minimum correlation length imposed to regularise the inversion) is increased, and the solution is improved where it is possible, and kept similar to the solution of the previous iteration at locations where there are not enough targets. Here, due to the limited spatial density of the targets and their inhomogeneous distribution in 3D space the minimum correlation length was 50~pc, much larger than the one achievable for dust (for the latter the available input catalogue of extinction and distance is an order of magnitude larger). 

\begin{figure*}   
*\centering 
\includegraphics[width=0.89\columnwidth]{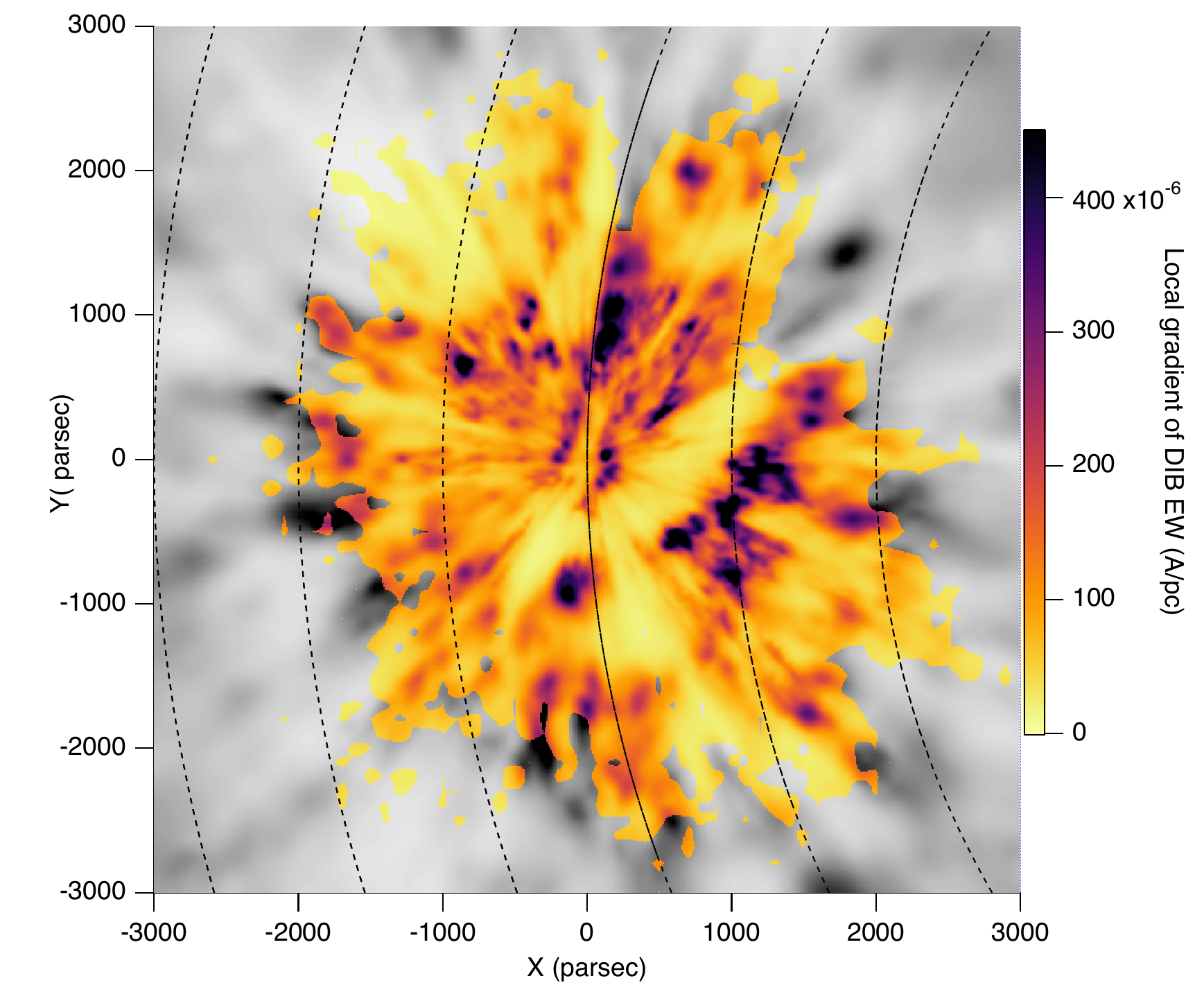}
\includegraphics[width=0.89\columnwidth]{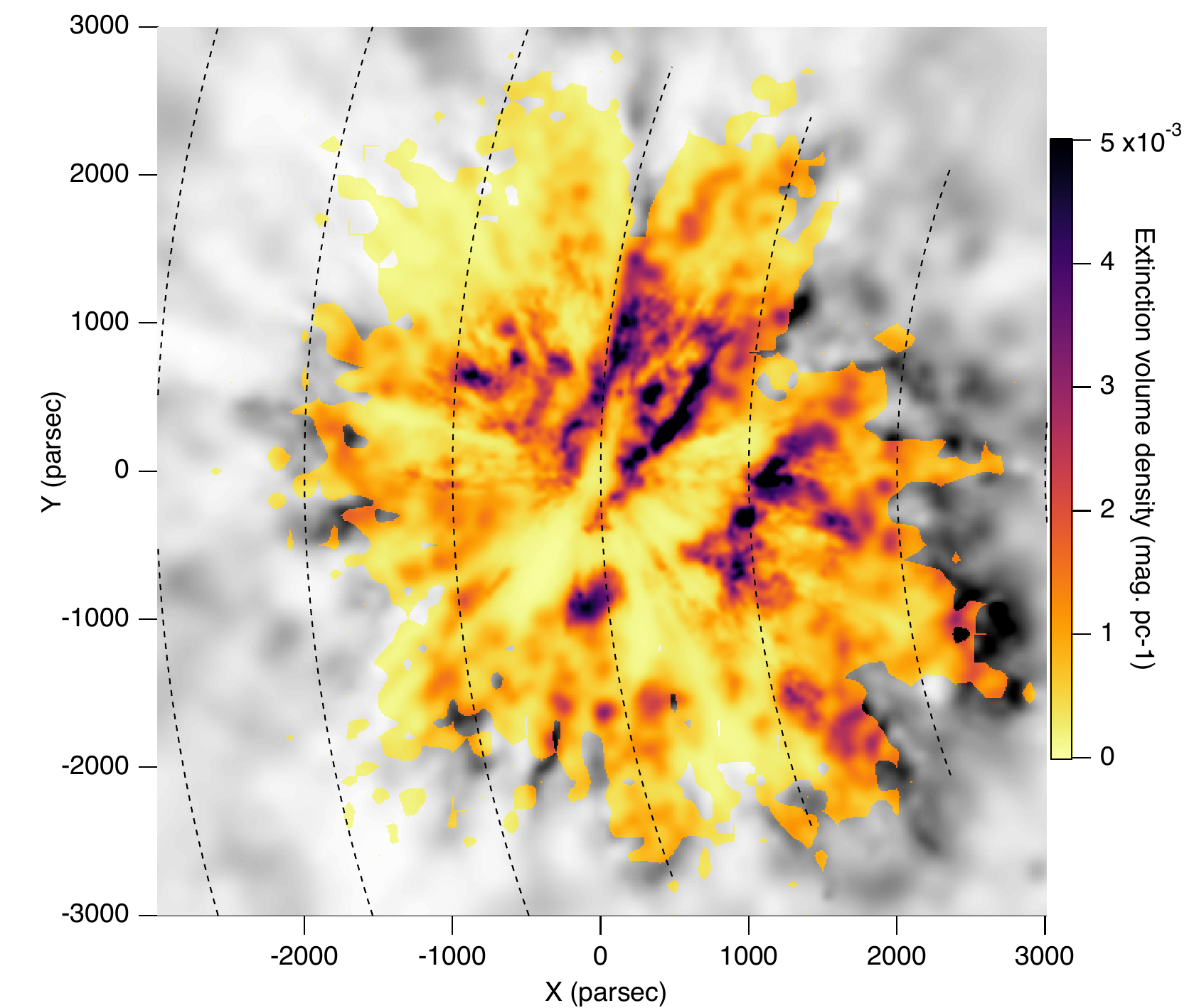}
\caption{862.1~nm DIB carrier volume density (left) and extinction volume density (right) in the Plane parallel to the Galactic plane and containing the Sun. Coordinates are in parsec from the Sun location. The Galactic Center is to the right and the longitude increases anti-clockwise. Dashed lines are iso-Galactocentric distances spaced by 1000pc. The central line corresponds to the Sun location and X=0. Black and white coded regions correspond to those with low target densities and more uncertain measurements.}
\label{fig:1}
\end{figure*}

\begin{figure*}
\centering 
\includegraphics[width=\columnwidth]{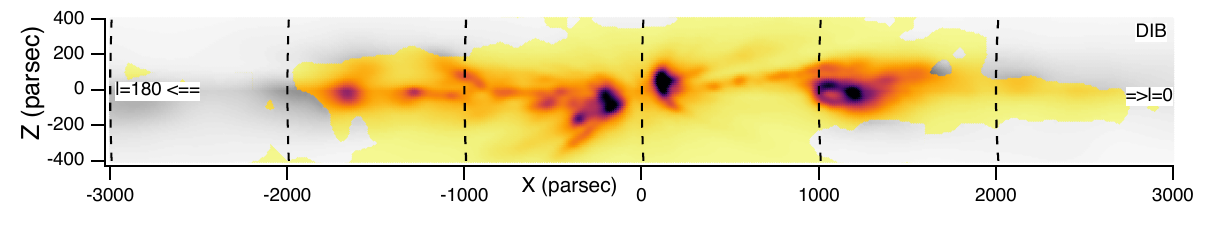}
\includegraphics[width=\columnwidth]{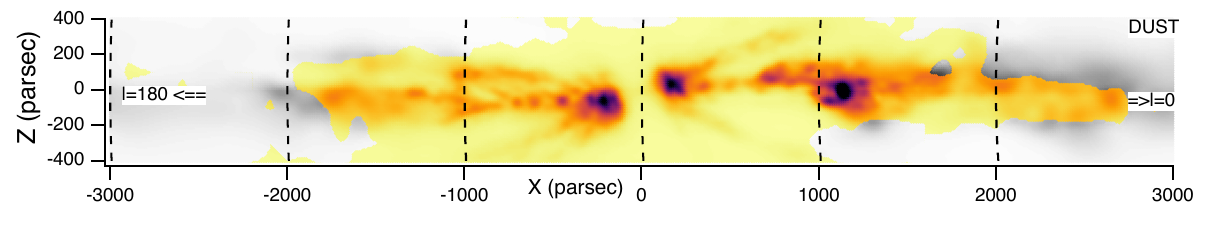}
\caption{Same as Fig.~\ref{fig:1} for a vertical plane containing the Sun and the Galactic centre direction (to the right). The North Galactic pole is to the top. (Colour codes are the same as in Fig.~\ref{fig:1}).}
\label{fig:2}
\end{figure*}

\subsection{Limits of the multi-scale Bayesian inversion of the DIB equivalent widths}

Figure~\ref{fig:5} shows the coverage (source density) of the input catalogue in the Galactic plane (all sources with abs(z) $<$ 50~pc are shown). The vertical scale upper limit is set such that all bins with at least 5 input sources are highlighted.

Beyond 2-3~kpc there are fewer objects (the exact value depending on the longitude and latitude), while the sampled volume increases exponentially. Including a margin, the reconstructed density cubes are cut off at 4~kpc. To quantify the fidelity of the 3D maps we also compute a ``maximum reliable distance'' map. We set the limiting distance to the distance of the most distant distance bin with at least five stars. Beyond this bin the accuracy of the map is severely hampered by the lack of input data. The maximum reliable distance map (as a function of galactic longitude and latitude) for the DIB reconstruction is shown in Fig.~\ref{fig:6} and is crucial in interpreting the 3D distribution of DIBs at the furthest distances. DIB densities near or beyond this distance are inaccurate and should be used or interpreted with great caution.

\begin{figure*}[h!]
\includegraphics[width=0.89\columnwidth]{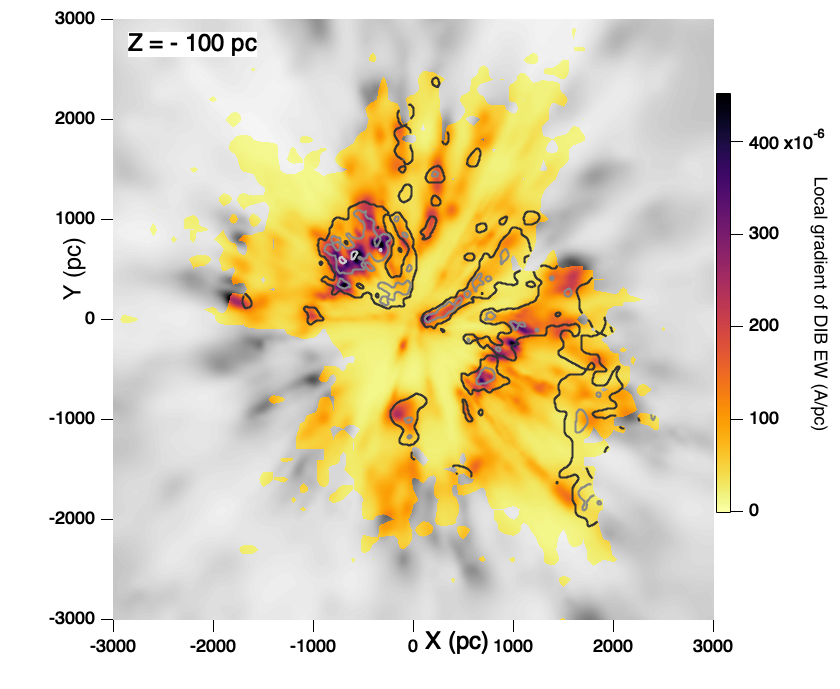}
\includegraphics[width=0.89\columnwidth]{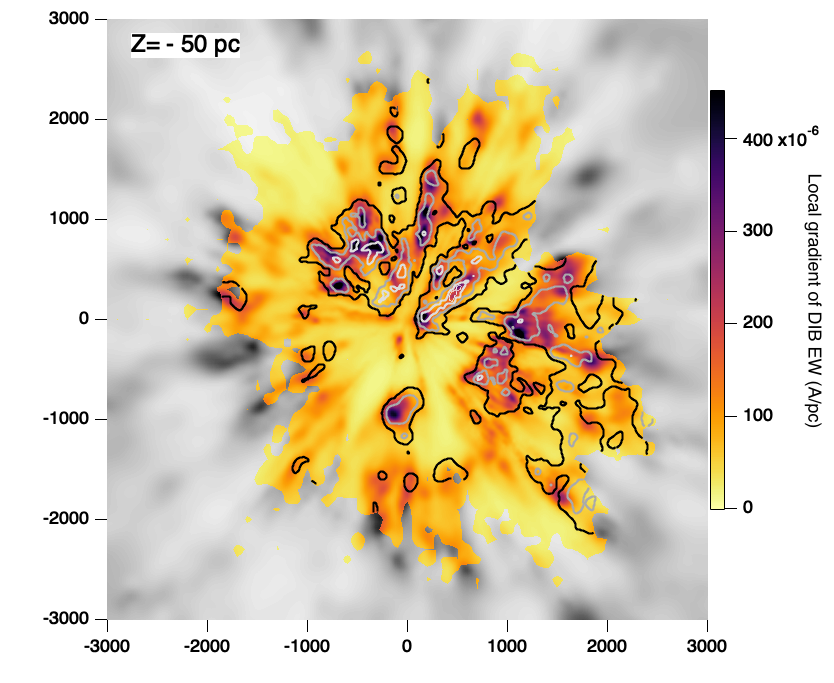}
\includegraphics[width=0.89\columnwidth]{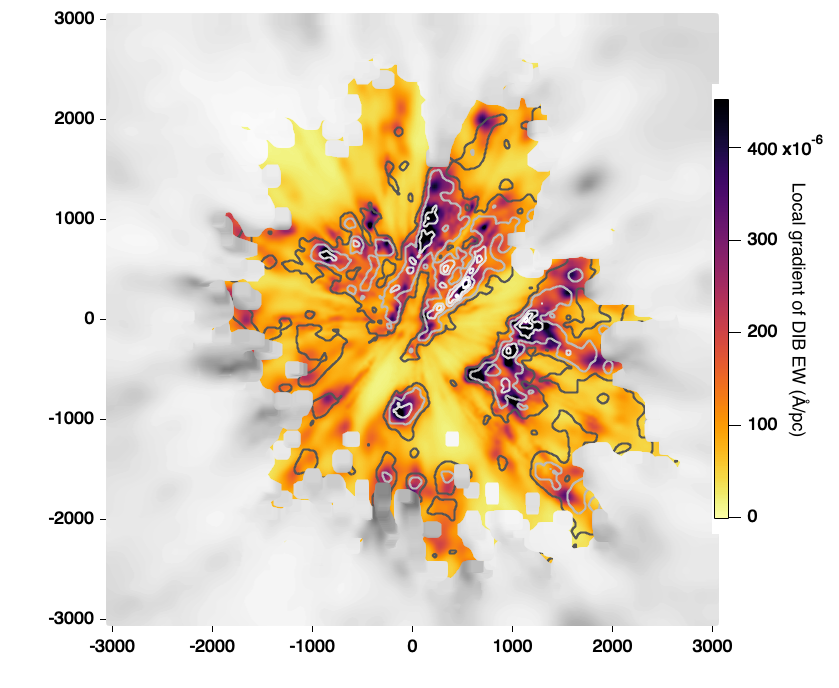}
\includegraphics[width=0.89\columnwidth]{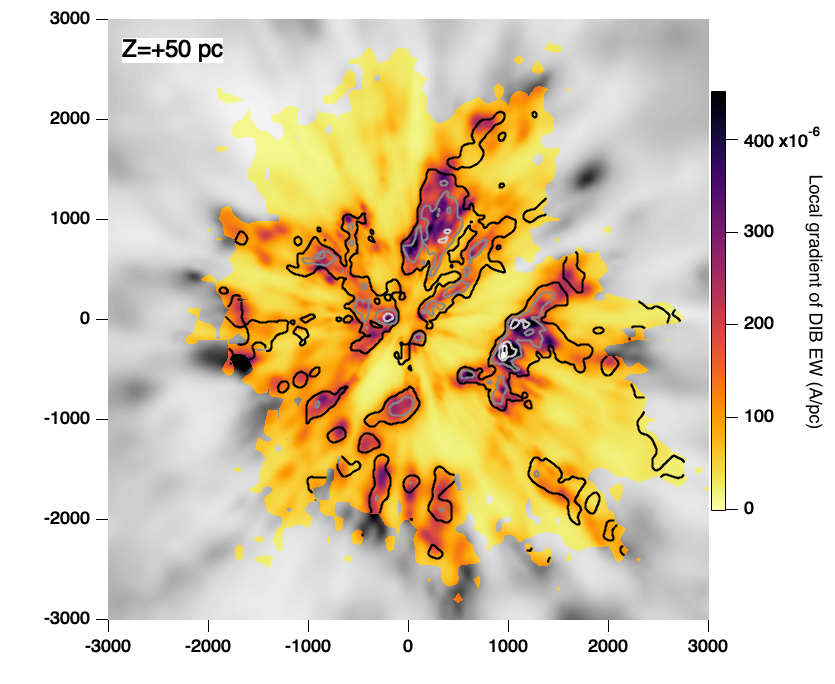}
\includegraphics[width=0.89\columnwidth]{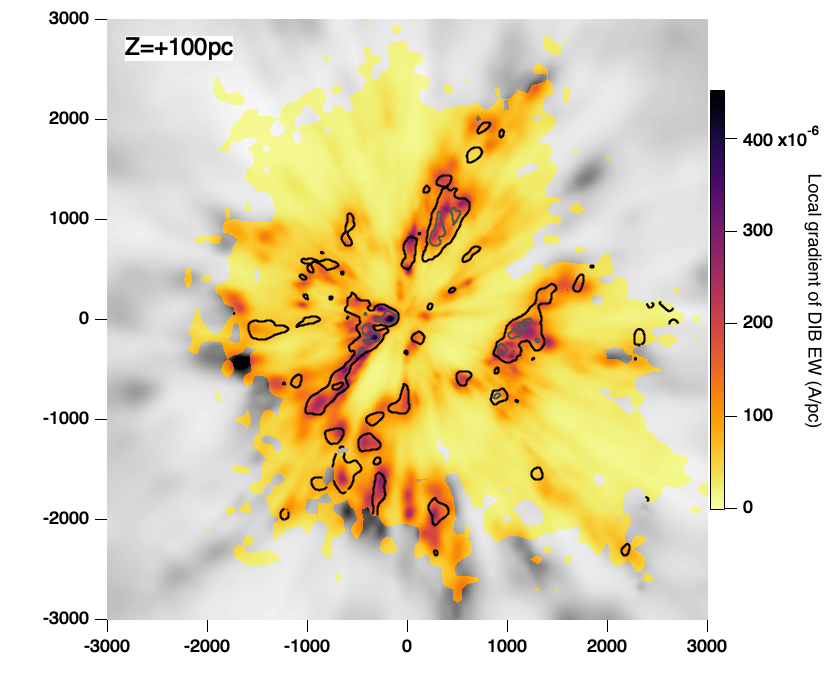}
\caption{862.1 nm DIB carrier volume density in planar surfaces parallel to the Galactic disk, and at different heights: -100, -50, 0, +50 and +100~pc. Superimposed are extinction density iso-contours in the same surfaces for densities 1, 2, 4, 6, 8 $\cdot$ 10$^{-3}$ mag~pc$^{-1}$.}
\label{fig:horizontal}
\end{figure*}

\subsection{Extinction density maps adapted to comparisons with the DIB maps}

To obtain a realistic comparison with the dust volume density, we did not make use of the extinction maps at higher spatial resolution that were based on much more numerous targets. Instead, we wanted to use a dust map of similar resolution  everywhere in 3D space and similar spatial extent. This is why, for such a direct comparison, we reconstructed extinction density maps using the same input sources as those selected for the DIB cubes. To do so, we started with the precise extinction maps from \citet{2022A&A...661A.147L} and, for each of the target stars used for the DIB map, we computed the cumulative extinction between the Sun and the star. This provided a reduced and inhomogeneous extinction dataset similar to the DIB dataset. We then performed the inversion of this dataset in the same way as for the DIBs.

Although the input catalogues for both DIB and extinction density are the same, and the general algorithm is the same, some details differ between the two datasets. The tomography algorithm applies a filtering direction by direction which depends on the consistency of the data along each direction. During the inversion we perform a $\chi^2$ test for the consistency in each direction -- we start by taking averages direction by direction with resolutions of 400~pc, 200~pc, and then going down to 100~pc, and finally 50~pc (the current limit with the available input data). If the $\chi^2$ is too large for a given direction this direction is deleted. Using the same $\chi^2$ threshold for both the extinction and DIB inversion inadvertently removes 3/4 of the directions for the latter. Thus, the threshold is relaxed for the DIB inversion.

Concerning the a priori DIB and extinction densities used during the tomographic processing, we used for both a prior distribution depending only on the distance to the Plane, i.e., independent of the distance to the Galactic Center. The same prior scale height is set to 200~pc for DIBs and dust. Independence with respect to the Galactocentric distance is important, it implies that the observed centre-edge variation of the DIB to dust ratio does not come from the a priori information injected into the tomographic process.

\begin{figure*}
\centering 
\includegraphics[width=\columnwidth]{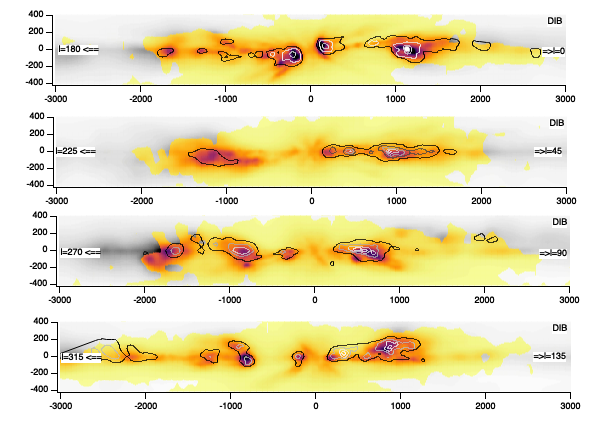}
\caption{Same as fig. \ref{fig:horizontal} for planar surfaces perpendicular to the Galactic disk and containing the Sun, oriented along various longitudes: 0--180\degr, 45--225\degr, 90--270\degr and 135--315\degr. Superimposed are extinction density iso-contours in the same surfaces for densities 1,2,4,6,8 $\cdot\ 10^{-3}$ mag~pc$^{-1}$.}
\label{fig:vertical}
\end{figure*}

\begin{figure*}
\centering 
\includegraphics[width=\columnwidth]{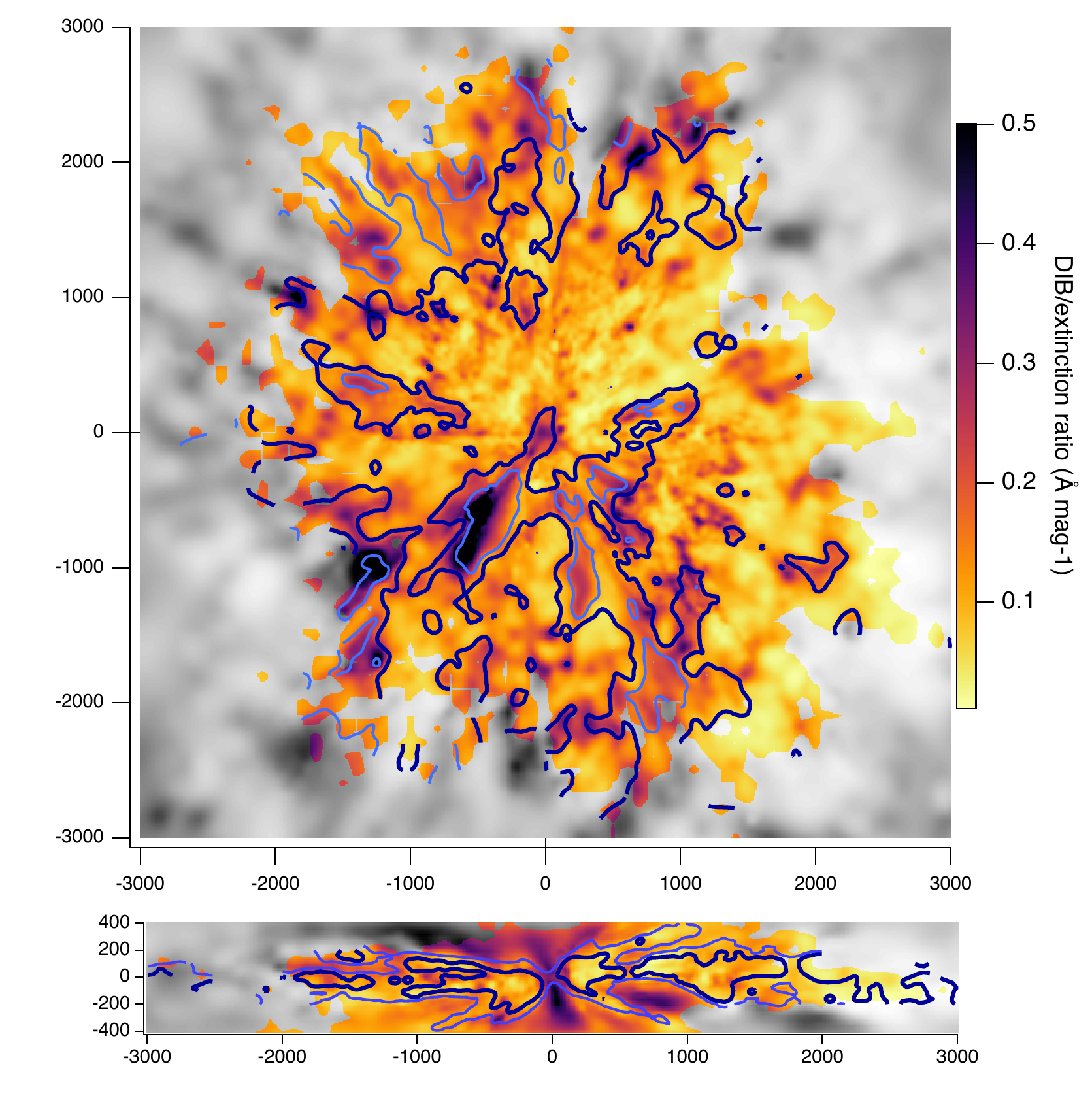}
\caption{Local values of the ratio between the DIB carrier volume density as measured by the EW local spatial gradient (in units of \AA~pc$^{-1}$) and the extinction volume density (in units of mag~pc$^{-1}$). The two DIB-to-dust ratio maps are for the same planes as in Figs.~\ref{fig:1} and~\ref{fig:2}. Iso-contours of extinction density for two low values (10$^{-5}$ (black) and 5 10$^{-4}$ (violet) mag~pc$^{-1}$) are superimposed.}
\label{fig:3}
\end{figure*}

\section{Results}\label{sec:results}

\subsection{DIB carrier and dust volume densities}

Figs.~\ref{fig:1} and~\ref{fig:2} show the inverted volume density of DIB carriers in the plane parallel to the Galactic disk and containing the Sun and in the meridian plane perpendicular to the disk, i.e. containing the Sun and oriented along the longitudes 0 and 180\degr.
Additional planar surfaces parallel to the Galactic disk (at different heights) and planes perpendicular to the disk oriented at various longitudes are shown in Figs.~\ref{fig:horizontal} and~\ref{fig:vertical}. Regions where the target density was too poor to perform the first inversion are colour-coded differently.

The resulting dust volume density is shown in the same planes as for the DIB carrier volume density in Figs.~\ref{fig:1} and~\ref{fig:2}. In order to allow an easier comparison between DIB and dust, Figs.~\ref{fig:horizontal} and~\ref{fig:vertical} display the extinction density iso-contours superimposed on the DIB carrier density for all shown planar surfaces.

The visual comparison between the DIB and dust distributions using these figures reveals a very strong similarity, i.e., regions with high dust density are characterised by a high DIB carrier density, and, reciprocally, regions with very low dust density are also characterised by very low DIB density. 
Such a similarity in shape is not very surprising, given the general correlation between DIB and dust. However, it is the first time it is observed in 3D on different distance scales. There are some differences in the locations of DIB and dust maxima locations, however distances at play are generally below or on the order of the map resolution.

More interesting are the trends of the density ratios. Despite the similarity of the shapes of dense structures or cavities, there are obvious significant differences in the relative levels of the DIB and extinction density, as shown in Fig.~\ref{fig:3} which displays for the same planes the ratio between the local volume densities of dust and molecules. The most visible differences are higher than average DIB-to-dust ratios in areas with low dust extinction, particularly in the extended cavity extending in the third quadrant, which is characterised by a very weak extinction and is filled with ionized gas \citep{2015JPhCS.577a2016L}. Similarly, while the DIB density is extremely low in the Local Chimney, following closely the dust cavity, (see the vertical plane image in Fig. ~\ref{fig:3}),  its ratio to the extinction density is above the mean, particularly below the Plane, where there is a high gas to dust ratio \citep{2015JPhCS.577a2016L}.


\subsection{DIB carrier and dust scale-heights or vertical distributions}

The 3D maps allow us to investigate the average out-of-plane dispersion (perpendicular to the disk) in more direct ways than using integrated quantities. A first approach is to adjust an exponential decrease with the distance to the Plane abs(Z) at all  locations in the Plane (X,Y-pairs). However, the distributions of dust and DIB carriers are strongly inhomogeneous, and such a concept of scale height is somewhat inappropriate, since it requires that the density decreases on both sides of the Z~=~0~pc horizontal plane in a quasi-exponential way, which is not the case in most locations. Even if we may adapt to the actual, wavy distribution of interstellar matter around the Plane by fitting an exponential decrease on both sides of the maximum density location (Z~=~ Z$_\mathrm{max}$, with Z$_\mathrm{max}$ positive or negative), in many regions there are several dense clouds at different altitudes and there is no adjustable exponential function.

To estimate the average out-of-plane dispersion of the two distributions and determine the extent to which they differ, we carried out two other calculations based on the 3D maps. First, we averaged the DIB density, the extinction density as well as their local ratio in $\delta$Z~=~10~pc thick horizontal layers. We restricted to $-2000$ $<$ X $<$ $+$2000~pc and $-$2000 $<$ Y $<$ $+$2000~pc. We found that the average ratio is minimal at Z $\simeq$ $+$50~pc, and is increasing with the distance to the Plane (see Fig.~\ref{fig:RATIO_CTOO_VS_Z} discussed in Sect.~\ref{AGBs}). More specifically, the average ratio increases by $\simeq$ 20~\% from Z = $-$50 to $+$150~pc (100~pc on both sides of the minimum), by $\simeq$ 40~\% from Z = $-$100 to $+$200~pc (150~pc on both sides of the minimum) and by $\simeq$ 75~\% at abs(Z) = 300~pc. The latter increase should be taken with caution due to the low number of targets and the particularly strong inhomogeneity of the distributions at such altitudes. 

\begin{figure*}
\centering 
\includegraphics[clip, trim=1cm 3cm 4cm 0cm, width=0.9\textwidth]{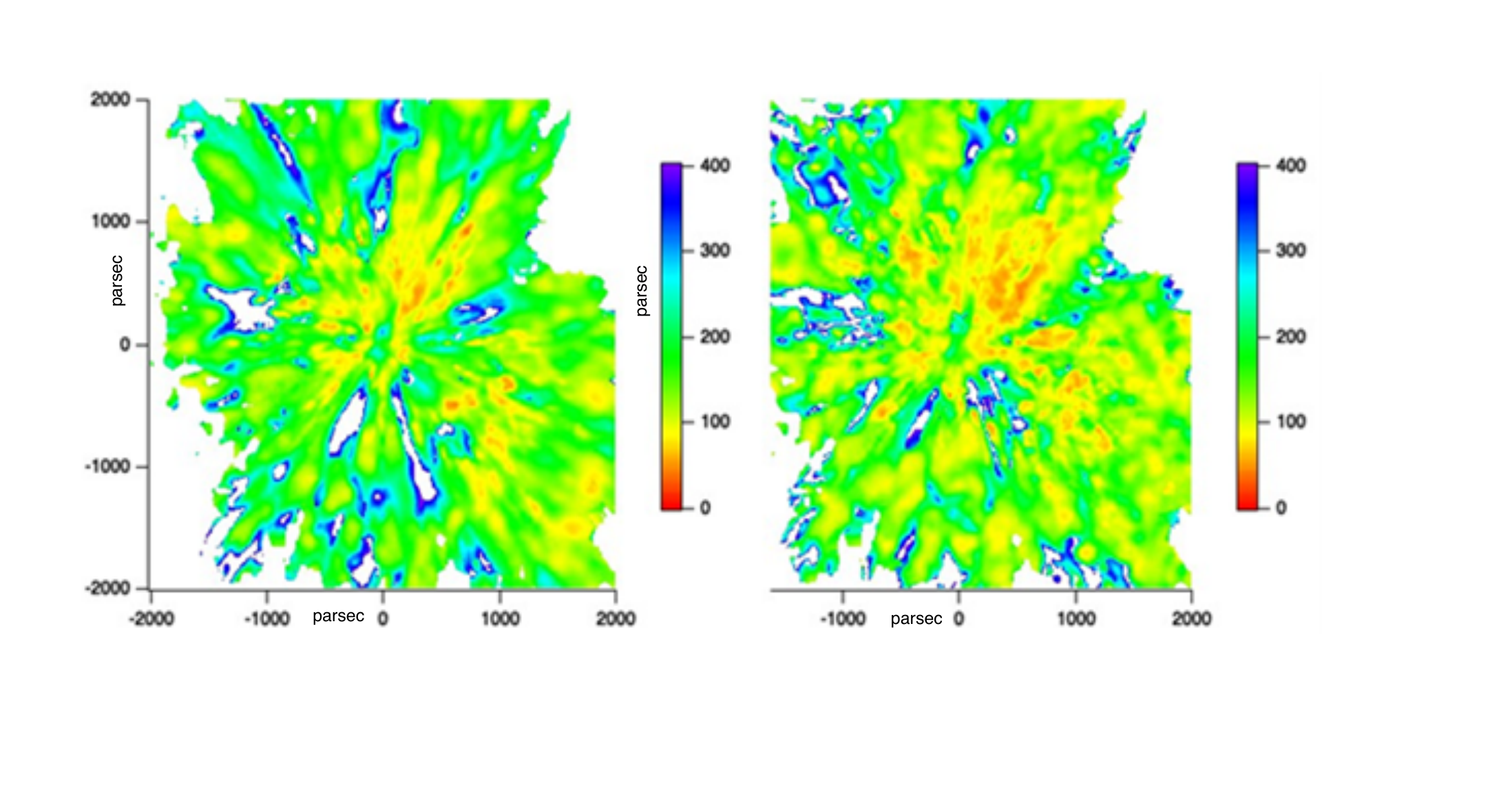}
\caption{Map of the DIB (left) and dust (right) fitted Gaussian half-widths of the out-of-Plane dispersion. Units are parsec. Such DIB and dust density scale-heights perpendicular to the Galactic (XY) Plane were derived by fitting a Gaussian function to the vertical (Z) distribution of DIBs and dust, respectively. The centroid of the Gaussian is fixed to Z$_\mathrm{max}$, the distance above/below the Galactic Plane with the highest dust density. See text for exclusion of spurious results.}
\label{fig:7}
\end{figure*}

The second way is as follows: for each (X, Y) pair we extracted from the map the dust and DIB density along the Z axis, computed separately for both quantities the altitude Z$_\mathrm{max}$ of the maximum density, then fitted a Gaussian to the Z-dependent distribution, centred on the altitude Z$_\mathrm{max}$. Although the distribution is very different from a Gaussian, such a procedure provides an order of magnitude of the dispersion on both sides of the Plane. We restricted the fits to locations (i.e. X,Y-pairs) where Z$_\mathrm{max}$ is comprised between $-$150~pc and $+$150~pc. Values outside this interval are signs of very low densities and uncertain measurements. After the fitting process, we also eliminated spurious results induced by lack of data or by too inhomogeneous distributions such as non-converging fits, Gaussian half-widths smaller than 50~pc which trace unique clouds, or on the contrary larger than 250~pc which correspond to absence of matter along Z (large regions devoid of interstellar matter). We also eliminated locations for which the altitudes of the two respective maxima Z$_\mathrm{max}$ (DIB) and Z$_\mathrm{max}$ (dust) were found to differ by more than 50~pc. Figure~\ref{fig:7} shows the widths found for DIB and extinction respectively as a function of coordinates X and Y. The mean ratio between the DIB width and the extinction width was found to be 1.14. The mean Gaussian half-width at 1/e was found to be 137~pc for the DIB and 123~pc for the dust. Note that only the ratio and the order of magnitude of the width are relevant here, since, as explained above, the Gaussian function is far from closely fitting the dispersion as a function of Z.

The relative increase in dispersion along Z we derive is smaller than the factor of two found by \citet{2014Sci...345..791K}, and in closer agreement with the results of \citet{2023A&A...674A..40G}. However, again, a strong variability of widths does exist depending on quadrants, hemispheres, with a larger DIB scale height in the Southern hemisphere. This may explain the results of \citet{2013ApJ...778...86K}, as the RAVE targets are predominantly at negative latitudes.


\subsection{DIB to dust ratio and Galactocentric distance}

One interesting result from the 3D mapping is the evolution of the DIB-to-dust ratio as a function of Galactocentric distance. 
Figure~\ref{fig:8} shows mean values of the DIB density, the extinction density, and their ratio, averaged in cylindric shells centred on the Galactic Center (GC), each of Galactocentric distance interval of 200pc.
Averages are restricted to areas that correspond to at least 5 targets in the inversion boxes. The size of the markers is increasing with the number of targets located in the shell. The correspondence between the size and this number is illustrated in the top graph. The ratio is increasing with the distance, i.e., increasing away from the GC, at an average rate of about 15~\% per~kpc.
The trend is about the same (results not shown here) if one restricts the vertical extent of the slices to 400~pc, and 200~pc, precluding an effect of scale height difference. This is the first time a trend of this kind is derived. 

\subsection{Comparisons with distributions and mass fluxes of AGB stars}\label{AGBs}

It is generally thought that most of the interstellar dust is produced by dying stars during the late Asymptotic Giant Branch (AGB) phases of their evolution. Two types of ejecta are found to coexist, namely carbon-rich and oxygen-rich dust particles, and new ground-based surveys as well as Gaia measurements and subsequent analyses now allow first estimates of the two productions. A recent catalogue of the nearby AGB locations and estimates of dust production has been produced by \citet{2022MNRAS.512.1091S}. We used the individual, estimated fluxes and locations for each target from this catalogue to compute the present C-rich and O-rich dust production in Galactocentric distance bins. The results and the total production ratios are displayed in Fig.~\ref{fig:8}. Although such catalogues are still limited in target numbers, resulting in a strong variability of the ratio in the outer Galaxy, trends can be detected, namely an increase of the O-rich dust with respect to the C-rich dust in the inner Galaxy and closer to the Galactic plane (or, conversely, an increase of C-rich dust relative to O-rich dust at Galactocentric distances larger than the one of the Sun). Such trends have already been noticed by \citet{2020ApJ...904...82Q,2020ApJ...892...52L,2022MNRAS.512.1091S}. Because DIB carriers require large quantities of carbon, while dust grains need oxygen to form the silicates, this may, at least partially, explain the radial gradient and the larger dispersion of DIB molecules around the Plane with respect to the extinction we find in the maps. 

\begin{figure}
\centering 
\includegraphics[width=0.9\columnwidth]{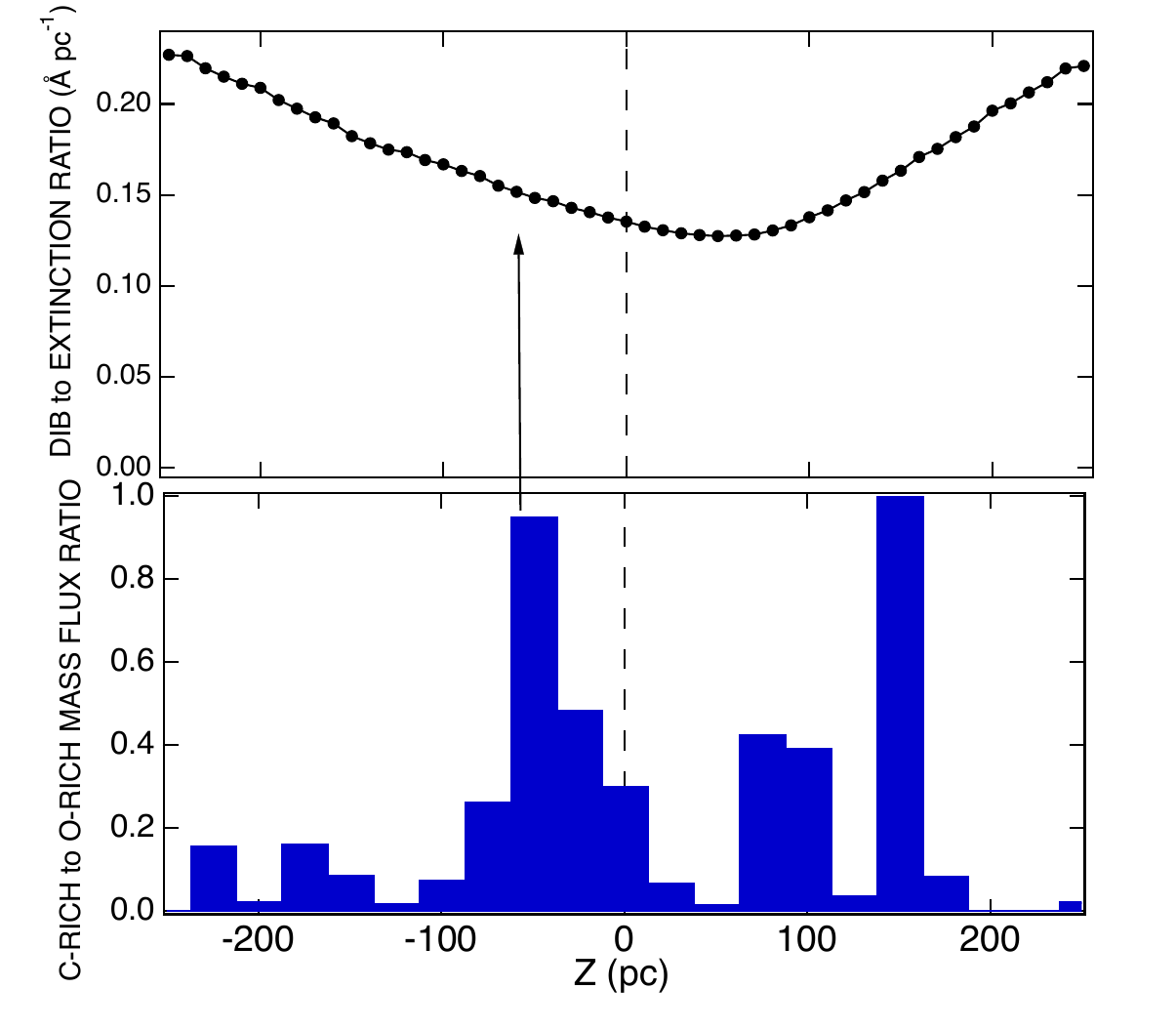}
\caption{Top: Average values of of the DIB to extinction density ratio in $\delta$Z = 10~pc horizontal slices as a function of the altitude Z.  Bottom: ratio between C-rich and O-rich dust fluxes from AGBs, computed based on the catalogue of \cite{2022MNRAS.512.1091S} for stars located in 25~pc wide horizontal slices.}
\label{fig:RATIO_CTOO_VS_Z}
\end{figure}

Given this potential link between the Galactocentric gradient of the DIB to dust ratio and the AGB distribution, we also computed the C-rich to O-rich dust flux ratio as a function of Z in 25~pc wide vertical slices, using the same catalogue by \citet{2022MNRAS.512.1091S}. The result is displayed along with the DIB to dust ratio computed in the same volumes in Fig.~\ref{fig:RATIO_CTOO_VS_Z}. There is a larger C/O ratio below the plane (from Z = $-$150 to Z = 0~pc) than above (between Z = 0 and Z = $+$150~pc), with a maximum at Z $\simeq$ $-$50~pc. In case this preponderance of C-rich dust at negative latitudes has an effect on the DIB density, this will result in an increase of the DIB to dust ratio below the Plane and, in turn, in a shift of its minimum towards positive altitudes, as is observed (the minimum is found at $+$50~pc). 
Finally, from the above works it is also inferred that the vertical distribution of AGBs is different for C-rich and O-rich types, with a larger concentration of O-type stars along the Plane and a bi-modal type distribution on both sides of the Plane for C-rich types. This may explain, at least partly, the increased density of DIB carriers, compared to dust grains, at larger distance from the Plane.

\section{Discussion and summary}\label{sec:discussion}

We have used the recently published catalogue of 862~nm DIB equivalent width measurements for individual stars, as well as corresponding Gaia positions and distances, to produce a 3D map of the DIB carrier density, using a hierarchical Bayesian inversion technique. Based on previous extinction maps, we have computed the extinction towards the same series of target stars constituting the DIB catalogue, and performed the same inversion on the resulting extinction catalogue. Doing so, we produced a 3D map of the extinction density at the same resolution at every location than for the 3D DIB map. This ensures pertinent comparisons between DIB carriers and dust volume densities.  

The visual comparison between the 3D maps of DIB and dust reveals a very strong similarity, i.e., regions filled with dust responsible for strong (resp. weak) optical extinction are characterised by a high (resp. low)  DIB carrier density. This is not surprising, given the correlations between columns of DIB carriers and extinction. However, it is the first time it is demonstrated in full 3D. On the other hand, there are significant differences in the relative levels of the DIB and extinction volume density, in the sense that higher than average DIB-to-dust ratios are found in areas with low dust extinction. The most conspicuous case is the wide, low dust cavity extending in the third quadrant, which is known to be filled with warm ionized gas, and where the average DIB to dust ratio is more than five times above the mean. There is a similar trend in the Local Chimney, especially below the Plane, where the gas to dust ratio is known to be especially low. Such a predominance of DIB carriers with respect to dust in low-extinction areas, by comparison with other regions, cannot be linked to the “skin effect” because the resolution of the map is too poor to reveal it, and the areas with high DIB-to-dust ratio are much wider than the thin regions of intermediate opacity surrounding clouds and giving rise to this effect. It must be linked to other properties of the DIB carrier formation and destruction. 

The full-3D maps allow us to investigate the average out-of-plane dispersion (perpendicular to the disk) in a more direct way than using integrated quantities. We averaged both DIB and dust densities in horizontal slices of various extent and examined how these quantities and their ratio vary as a function of the distance to the Galactic plane. We found that the DIB carriers generally extend farther away from the Plane than the dust grains producing the bulk of the extinction, however, the ratios vary strongly from one location in the Plane to the other and reflect the inhomogeneity of both distributions. The average ratio is minimal at Z $\simeq$ $+$50~pc, increases by $\simeq$ 20~\% from Z = $-$50 to $+$150~pc (100~pc on both sides of the minimum), by $\simeq$ 40~\% from Z = $-$100 to $+$200pc (150~pc on both sides of the minimum). In a second approach, we fitted Gaussian distributions along the vertical axis Z for each (X, Y) position in the Plane. Doing so, we derived DIB and dust Gaussian half-widths (at 1/e), H$_\mathrm{dust}$ = 127~pc and H$_\mathrm{DIB}$ = 133~pc. This again shows that the average dispersion around the Plane is larger for the molecules than for the dust grains, with an average relative difference of 11~\% for the widths. This is less than the factor of two found by \citet{2014Sci...345..791K}, and in closer agreement with \citet{2023A&A...674A..40G}. However, a strong variability of widths does exist depending on areas, with a larger scale height in the Southern hemisphere. This may explain the results of \citet{2013ApJ...778...86K}, as the RAVE targets are predominantly at negative latitudes.

\begin{figure}
\centering 
\includegraphics[width=0.8\columnwidth]{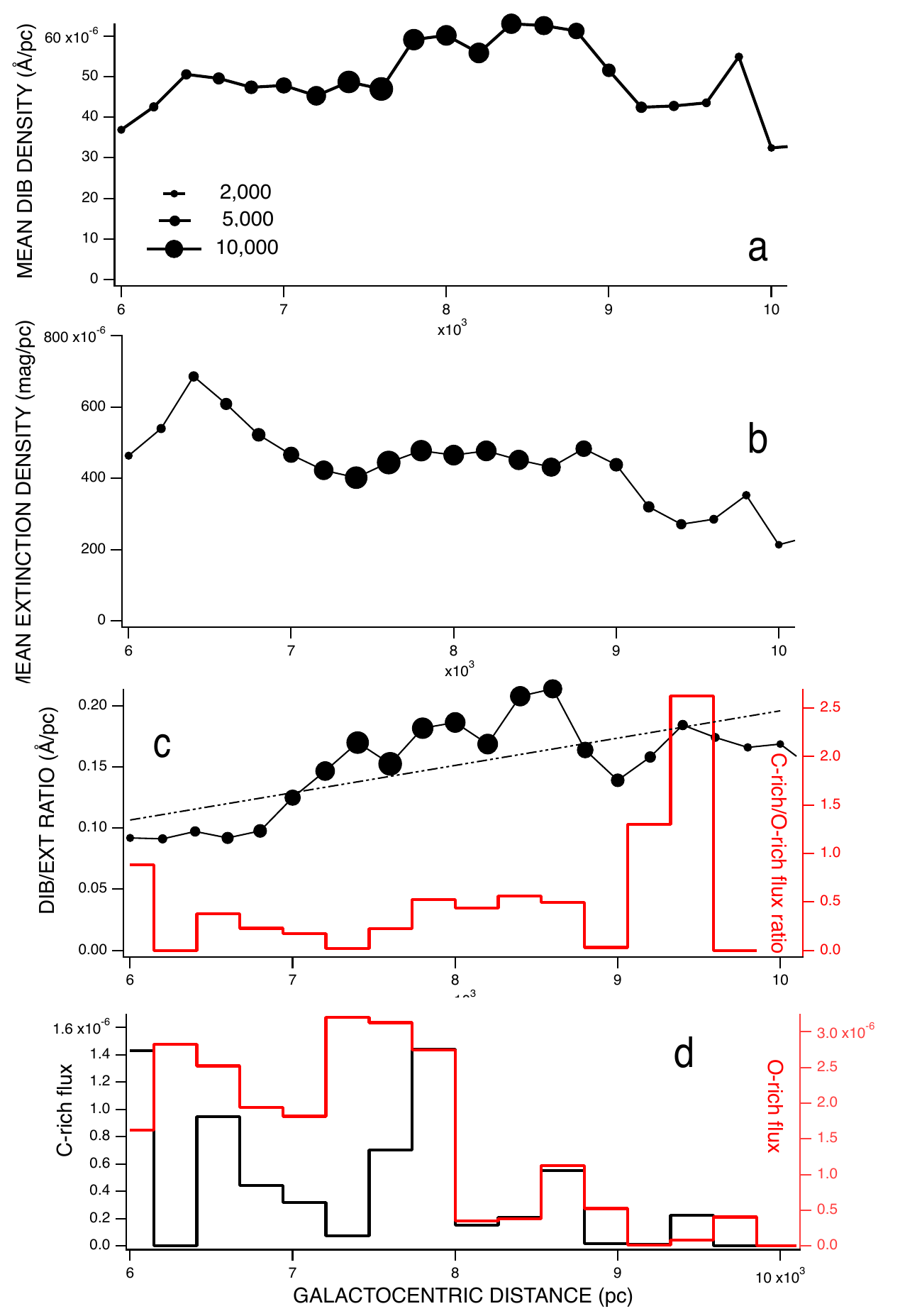}
\caption{Average values of DIB (a) and extinction (b) volume density in 100~pc thick cylindric shells for varying Galacto-centric distances, and their ratio (c). All cells of the 3D maps were averaged except for low density targets (back contour in Fig. \ref{fig:5}. The size of the markers is increasing with the number of targets located in the shell. The correspondence between the size and this number is illustrated in the top graph. A linear fit to the ratio is drawn in panel c. Also shown in c is the ratio between C-rich and O-rich total fluxes from AGBs located in 265~pc thick cylindric shells, computed based on the catalogue of \cite{2022MNRAS.512.1091S}. The corresponding C- and O-rich total fluxes are displayed in panel d.}
\label{fig:8}
\end{figure}

Finally, we averaged densities and ratios in cylindric shells whose axes are vertical and contain the GC, for various Galactocentric distances, and found that the DIB to dust ratio is increasing radially (see Fig.~\ref{fig:8}). The average gradient is around 15~\% per~kpc in the Sun vicinity. The potential explanation for this new result can not be linked to the radial gradient of metallicity, which, according to Magellanic clouds studies is expected to produce an effect opposite to what is observed.

Although our conclusions are limited by the resolution of the maps, the observed trends deserve attention. The first result is the apparently generalized increase of the 860.2~nm DIB to extinction ratio in low dust areas (see Fig.~\ref{fig:3}). If the macro-molecule at the origin of the DIB is one of the numerous products of the very rich chemistry occurring at the surface of icy grains in dense cloud cores, the results suggest that, once they have been released from the grains in the gas due to shock or radiation, these molecules survive during a period of time long enough to be detected away from the clouds in low density regions. This implies a strong resilience of the 860.2~nm DIB carriers at locations where dust grains are partly destroyed. If 860.2~nm DIB carriers are instead built gradually in stellar ejecta and in the diffuse interstellar medium, they must be preferentially associated with grains smaller than those responsible for the extinction in the visible. The same types of conclusions arise from the DIB out-of-plane dispersion, found to be in average higher than the dust dispersion. DIB carriers produced in dense areas close to the Galactic plane may be released at higher altitude and maintained far from the dense clouds in lower density gas where grains are destroyed or settle down to the Plane by gravity. In both cases, dust-poor regions or regions above/below the disk, the observed trends imply that the DIB carrier density follows the gas density and not solely the dust concentration.

We investigated the potential role of the spatial distribution of AGB stars. Their distribution in the Milky Way is increasingly well measured thanks to Gaia. New observations distinguish the oxygen-rich (O-type) and carbon-rich (C-type) AGB stars and recent analyses from space and ground show that in the Sun vicinity O-rich AGB stars are predominant in the inner Galaxy and close to the Plane. C-rich dust is expected to favor the formation of the DIB carriers, while O-rich ejecta are needed to form silicate grains. We used the catalogue of \citet{2022MNRAS.512.1091S} and inferred from it that the production of C-rich dust is increasing radially by comparison with O-rich dust, in agreement with the direction of the DIB to dust gradient we observe. Interestingly, the same catalogue shows an increased production of C-rich dust below the Plane (over the first 100~pc), which could have a link with the observed location of the minimum DIB to dust ratio above the Plane, shifted to Z = $+$50~pc. Finally, the bi-modal vertical distribution of C-rich AGB stars with maxima on both sides of the Plane, at variance with the maximal concentration of O-rich stars close to Z = 0~pc, may potentially, at least partly, explain the larger dispersion around the Plane of DIBs with respect to grains. 

While still limited by the map resolution and to be confirmed, our results on a potential link with C-rich AGB stars show that full-3D mapping of DIB absorption data opens the way to new diagnostics about the processes involving their carriers. Ongoing and future massive spectroscopic surveys will allow the application of the 3D mapping technique to stronger and narrower DIBs to produce more resolved maps, test differences among the various DIBs, while parallel improvements of stellar catalogues should help confirm, or not, the link with the various types of stars and environments.

\begin{acknowledgements}
This project has received funding from the European Union’s Horizon 2020 research and innovation program under grant agreement No 101004214 (EXPLORE project – https://explore-platform.eu).

This article/publication is based upon work from COST Action CA21126 - Carbon molecular nanostructures in space (NanoSpace), supported by COST (European Cooperation in Science and Technology)."

We made use of data from the European Space Agency (ESA) mission Gaia (https://cosmos.esa.int/gaia), processed by the Gaia Data Processing and Analysis Consortium (DPAC, https://cosmos.esa.int/web/gaia/dpac/consortium). Funding for the DPAC has been provided by national institutions, in particular, the institutions participating in the Gaia Multilateral Agreement.

This research has made use of the VizieR catalogue access tool, CDS, Strasbourg, France (DOI: 10.26093/cds/vizier). The original description of the VizieR service is published in 2000, A\&AS 143, 23.

\end{acknowledgements}

\bibliographystyle{aa} 
\bibliography{3DDIBS.bib} 


\appendix
~
\section{Additional figures}

\begin{figure*}[h!]
\centering 
\includegraphics[width=0.49\columnwidth]{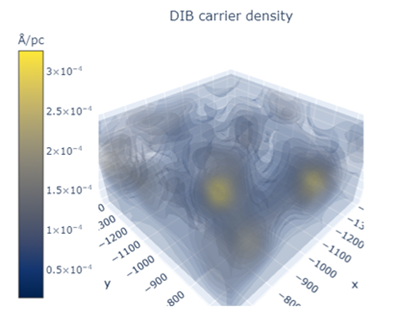}
\includegraphics[width=0.49\columnwidth]{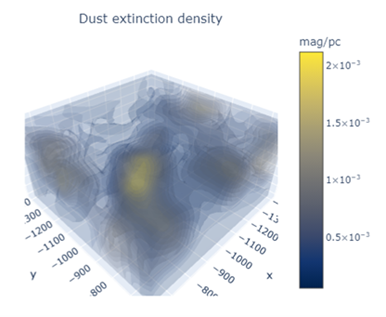}
\caption{Volume rendering of the diffuse interstellar band carrier (left) and dust (right) for a sub-region of the full 3D data. The centre of the 3d volume plot is X = $-$1000~pc, Y = $-$1000 pc, Z = $-$102.5~pc, with size of 400~pc in X and Y, and 200~pc in Z. Density surfaces colour coded. The full cube can be explored by downloading from Zenodo or viewed using the interactive tool G-Tomo on \url{https://explore-platform.eu}.}
\label{fig:3D}
\end{figure*}

\end{document}